\documentstyle[sprocl,epsfig]{article}

\def\be{\begin{equation}}
\def\ee{\end{equation}}
\def\bea{\begin{eqnarray}}
\def\eea{\end{eqnarray}}

\begin{document}

\title{THE STATUS OF CHARMONIUM PRODUCTION IN PHOTON-PHOTON COLLISIONS}

\author{Cong-Feng Qiao\footnote{JSPS Research Fellow}}

\address{Department of Physics, Hiroshima University \\
Higashi-Hiroshima 739, Japan
\\E-mail: qiao@theo.phys.sci.hiroshima-u.ac.jp} 


\maketitle

\abstracts
{In this talk I briefly review the status of Charmonium
production in photon-photon collisions. I would like to mention that
although the preliminary data were obtained in experiment,
theoretical calculation is not in a compatible status, that is the
result of dominant contributing process is still not available.}

\section{A Brief Overview of Onium Production Theory and Models}

Quarkonium is a bound state of heavy quark and its antiquark
mediated by the strong interaction (QCD).
The quarkonia can be classified and labeled in the conventional 
spectroscopic way $n$ $^{2S + 1}L_J (J^{PC})$, 
where $n$ is the radial quantum number; $P$ is the parity and $C$
is the charge conjugation; 
S, L, and J are total intrinsic spin, orbital angular momentum, and
total angular momentum, respectively.

Since the discovery of $J/\psi$ in 1974, quarkonium physics has 
become one of the most fruitful areas in high energy physics. 
With the development of experiment in the past two decades
a great advance has been made in understanding of the nature 
of quarkonium production and decays. 
At the time being, the wisdoms on quarkonium production and decays
can be categorized as following theory and models:

\vskip 1mm
\noindent
{\bf $\bullet$ Color-Singlet Model \cite{t.a.degrand}}

In color-singlet model, quarkonia are interpreted as
non-relativistic bound states of $Q\bar{Q}$ pair.
It assumes that the heavy quark pair 
produced in high energy collision can bind to form a given 
quarkonium state if the $Q\bar{Q}$ is created  
with exactly the same quantum numbers possessed by the 
bound state. And, the quarkonium production and decay
amplitudes are supposed to be factorized into short- 
and long-distance sectors. The former 
can be calculated by using perturbative QCD; the latter, 
referring to non-perturbative effect, can be 
absorbed into a wave function factor; i.e.,
$$d\sigma(\psi + {\sc{x}}) = d\sigma (c\bar{c_1}(^3S_1) + \sc{x})
|R_\psi(0)|^2.$$
The wave function can be either determined phenomenologically
from experimental measurements of quarkonium leptonic decays
or calculated from bound state potential model.

\vskip 1mm
\noindent
{\bf $\bullet$ Color Evaporation Model \cite{h.fritzsch}}

An alternative prescription for quarkonium production is the so-called
color-evaporation(duality) model. In this approach, the probability
of a $Q\bar{Q}$ pair with invariant
mass between $2m_c$ and $D\bar{D}$ threshold $2m_D$(in case of
Charmonium) evolving into a quarkonium state will be taken nearly
independent of its color and spin states.
e.g., for $J/\psi$ production, the cross section can be written as 
$$\sigma(J/\psi) = \hat{\sigma} (c\bar{c}(4m_c^2<s<4m_D^2)) f_{J/\psi},$$
where $\hat{\sigma}$ is the cross section for producing a $c\bar{c}$ 
pair with invariant mass below $D\bar{D}$ threshold, $f_{J/\psi}$ 
is a phenomenological parameter. The cross section 
$\hat{\sigma}(c\bar{c})$ is spin-summed and can be in both color-singlet
and -octet configurations. This model has the flaw of incapable of 
describing the variation of production ratios for different states, 
though with some phenomenological success.

\vskip 1mm
\noindent
{\bf $\bullet$ NRQCD Factorization Theory \cite{g.t.bodwin}}

Non-relativistic QCD(NRQCD)
provides a rigorous QCD analysis of the production 
and decays of heavy quarkonium, which enables one to make  
perturbative corrections to all orders in $\alpha_s$, and relativistic 
corrections as well. The key point of this novel theory is that 
the inventors noticed that in quarkonium production and decays several 
typical energy scales are well-separated,
$$(M_Q v^2)^2 \ll (M_Q v)^2 \ll M_Q^2.$$
With this hierarchy, the NRQCD effective Lagrangian can be 
expressed as
$${\cal L}_{\rm NRQCD} = {\cal L}_{light} + {\cal L}_{heavy} + 
\delta{\cal L},$$
where $ {\cal L}_{light} + {\cal L}_{heavy}$ describes 
ordinary QCD coupled to a Schr\"odinger field theory for the heavy 
quarks and antiquarks. The relativistic effects of full QCD are 
reproduced through the correction term $\delta{\cal L}$ in the 
Lagrangian. 

In NRQCD formalism, the inclusive production cross-section of heavy 
quarkonium is argued taking a factorized form 
$$d\sigma(H + X) = \sum d\hat{\sigma}(c\bar{c} + X)<{\cal{O}}_H> .$$
Here, $d\hat{\sigma}(c\bar{c} + X)$
is the hard part calculable using perturbative QCD, $<{\cal{O}}_H> $
is the non-perturbative sector which can be expressed as vacuum matrix
elements of NRQCD four quark operators. In the NRQCD Lagrangian, 
four-fermion operators can couple to both color-neutral states and
colored states, which makes the NRQCD distinctively different
from the color-singlet hypothesis in describing the quarkonium
production mechanism, i.e., the Octet mechanism may play a role
in quarkonium production as well. 

\section{Quarkonium Production in Photon-Photon Collisions}

The NRQCD has a list of merits in describing heavy quarkonium
production and decays, especially in properly regulating the 
singularities appeared in
color-singlet model and explaining the Tevatron large-$p_T$ $\psi'$
surplus problem based the color-octet model \cite{com}. However, 
it is not the end of story. There still lacks of direct evidence for 
the octet scenario at currently running colliders. Not to mention the 
large-$p_T$ $\psi'$ polarization "disaster" it encounters 
recently \cite{rothstein}. The point is that in principle 
NRQCD should be a correct theory in heavy quark limit, but in practice 
whether it can be applied to the Charmonium system is not clear.

People believe that non-hadronic collisions may give more clear 
signals and predictions by experiment and theory respectively than
hadronic ones, and think that the study of quarkonium production at 
linear colliders may be helpful in clarifying the Onium production 
mechanism. 
In recent years several new concepts on linear colliders aimed at
providing collisions at the center-of-mass energy from hundreds GeV 
to multi-TeV with high luminosity were proposed and the feasibilities 
pretested, such as JLC at KEK, TESLA at DESY and CLIC at CERN, 
etc. Theoretically, high-intensity photon beams may be obtained 
by the Compton backscattering of laser light off the linac electron 
beams, and photon-photon collision with approximately the 
same luminosity as that of the $e^+$ $e^-$ beams. Such a photon linear
collider can have high energy up to TeV order.
During the past more than twenty years researches on quarknoium 
production at $e^+$ $e^-$ colliders at various energies were 
carried out in detail \cite{ee1}. However, studies of photon-photon 
scattering are very limited and have just begun \cite{gg1}(here, 
we focus on the direct photon-photon collision, for the resolved 
case see, e.g., ref. 6), though the preliminary results were obtained 
from LEP II data. 

In $\gamma\gamma$ scattering, at leading order in $\alpha$ the 
$J/\psi$ is produced via the process
\begin{eqnarray}
  \label{eq:leading1}
  \gamma + \gamma \rightarrow J/\psi + \gamma\;.
\end{eqnarray}
However, since on the scale of heavy quark mass, the strong coupling 
constant is not too small, the process 
\begin{eqnarray}
  \label{eq:leading2}
  \gamma + \gamma \rightarrow J/\psi^{(8)} + g
\end{eqnarray}
may compete with the pure electromagnetic process (\ref{eq:leading1}) 
through the Color-Octet mechanism \cite{gg1}. Here, $J/\psi^{(8)}$ 
denotes those states evolved from the Color-Octet states. 

When going up one order in $\alpha_s$, one may still expect to 
obtain the same order of magnitude in the $J/\psi$ production rate, 
because in this case $J/\psi$ may be produced in color-singlet and 
therefore will compensate for $\alpha_s$ suppression from the 
non-perturbative sector relative to the octet process 
(\ref{eq:leading2}). This argument was confirmed recently by the 
calculation of double quarkonium, the $J/\psi$, production in direct 
photon-photon collision\cite{c.f.qiao}, which is a sub-category of the 
inclusive $J/\psi$ production process at order $\alpha^2\alpha^2_s$. 
That is the process
\begin{eqnarray}
  \label{eq:leading3}
  \gamma(k_1) + \gamma(k_2) \rightarrow J/\psi(P) + J/\psi(P')\;. 
\end{eqnarray}

\begin{figure}[tbh]
\begin{center}
\vskip -4cm
\epsfig{file=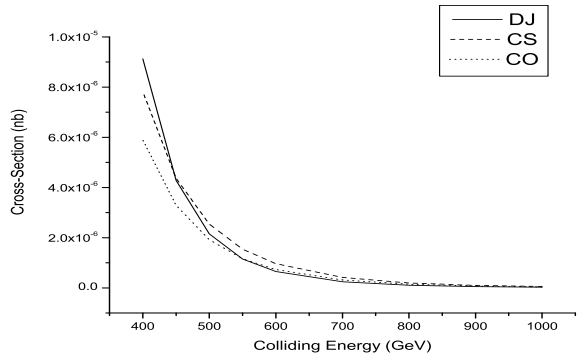,bbllx=90pt,bblly=300pt,bburx=230pt,
bbury=420pt,width=4cm,height=2.5cm,clip=0}
\end{center}
\vskip 6cm
\caption[bt]{Energy dependence of the cross-sections. DJ: the 
double $J/\psi$ process(\ref{eq:leading3}); CS: the color-singlet 
process (\ref{eq:leading1}); CO: the color-octet process 
(\ref{eq:leading2}).}
\label{graph2}
\end{figure}    

The results for processes (1) -- (3) are shown in Figure 1. 
Since projected linear colliders with a luminosity 
of hundreds of fb$^{-1}$ per year, and the 
integrated total cross-sections increase with the 
colliding energy decreasing, we may have 
hundreds of events being observed in one year at colliding energies 
500 GeV or less. It is obvious that the color-octet process is the 
smallest one over the entire energy scope of LEP II to 
next generation of linear colliders in the three
processes being concerned, though they are in the same order. 

In addition it was proved in Ref. 8 that 
the cross-section for single $J/\psi$ production 
via only the fragmentation mechanism would be about one order 
larger than that of processes (1) -- (3) at 500 GeV.  
That means that the process at order $\alpha^2\alpha_s^2$ for single 
$J/\psi$ inclusive production should be the dominant one in the
photon-photon collision, which unfortunately is 
still not studied in theory.

\section{Concluding Remarks}

We have given a brief review of the previously investigated
works appeared in literature on $J/\psi$ production at photon 
colliders. It is found that at a moderate energy of next generation 
photon colliders, there could be hundreds of events to be detected 
per year with high projected luminosity. Since the production rates 
of $J/\psi$ via the color-octet mechanism, the electromagnetic process, 
and the double production are almost the same in the full scope of the 
colliding energy, to differentiate the color-octet mechanism from 
the color-singlet one in these processes, experimentally one 
should detect not only the $J/\psi$ but also other final states.
Furthermore, without distinguishing the final states in experiment
we can not make any conclusion on $J/\psi$ production mechanisms
from photon-photon collision, since the dominant production process, 
the inclusive one at order $\alpha^2\alpha^2_s$, is left 
un-investigated in theory.
\section*{Acknowledgments}
This work was supported by a Grant-in-Aid aid of JSPS committee.

\end{document}